\documentclass[sigconf]{acmart}
\AtBeginDocument{%
  }

\copyrightyear{2026}
\acmYear{2026}
\setcopyright{cc}
\setcctype{by}
\acmConference[SIGIR '26]{Proceedings of the 49th International ACM SIGIR Conference on Research and Development in Information Retrieval}{July 20--24, 2026}{Melbourne, VIC, Australia}
\acmBooktitle{Proceedings of the 49th International ACM SIGIR Conference on Research and Development in Information Retrieval (SIGIR '26), July 20--24, 2026, Melbourne, VIC, Australia}
\acmDOI{10.1145/3805712.3809871}
\acmISBN{979-8-4007-2599-9/2026/07}

\usepackage{graphicx}
\usepackage{multirow}
\usepackage{booktabs}
\usepackage{amsmath}

\usepackage{todonotes}

\usepackage[normalem]{ulem}

\newcommand{\crdel}[1]{{\color{red}\sout{#1}}}

\begin{document}

\title{Layer-wise Token Compression for Efficient Document Reranking}

\author{Shengyao Zhuang}
\email{syzhuang@amazon.com}
\affiliation{
  \institution{Amazon AGI}
  \city{Sunnyvale}
  \state{California}
  \country{USA}
}

\author{Zhichao Xu}
\email{xzhichao@amazon.com}
\affiliation{
  \institution{Amazon AWS}
  \city{Santa Clara}
  \state{California}
  \country{USA}
}

\author{Ivano Lauriola}
\email{lauivano@amazon.com}
\affiliation{
  \institution{Amazon AGI}
  \city{El Segundo}
  \state{California}
  \country{USA}
}

\begin{abstract}
Transformer-based document cross-encoder rerankers are a central component of modern information retrieval systems. 
Despite their success, these models suffer from high computational costs due to processing long query-document sequences at inference time. 
A known approach to improve efficiency is token compression, which consists of aggregating groups of tokens together in the initial embedding layer, reducing the effective number of tokens and making the computation faster. 
While token compression has proven to be successful for bi-encoder retrievers, we empirically observed that this approach may be ineffective for cross-encoder rerankers. 
In this paper, we propose \textbf{Layer-wise Token Compression (LTC)}, which applies adaptive token pooling at intermediate transformer layers. 
Through extensive ablation studies on MS MARCO passage and document ranking tasks, we demonstrate that compression at middle layers preserves ranking quality while increasing inference QPS by up to 25\% for passage ranking and up to 116\% for document ranking. We also extend LTC to listwise LLM rerankers and show that the same approach can be easily applied to long-context listwise reranking, where the QPS improvements are even greater. More surprisingly, when applying rerankers trained on short passages to long-document ranking tasks, models trained with compression outperform their uncompressed counterparts, suggesting that compression may act as a beneficial regularizer that encourages length-invariant representations.
\end{abstract}



\maketitle

\section{Introduction}

Transformer-based document reranking models have become essential in modern information retrieval systems, including Transformer-based cross-encoder models~\cite{nogueira2019passage,bert2021yates,xu-etal-2025-distillation,xu2025surveymodelarchitecturesinformation}, generative LLM-based listwise rerankers~\cite{sun2023chatgpt,ma2023zero,chen2025first,gangi-reddy-etal-2024-first} and, more recently, reasoning-based rerankers \cite{weller2025rank1testtimecomputereranking, zhuang2025rankr1enhancingreasoningllmbased, yang2025rankktesttimereasoninglistwise}. 
Although these methods achieve remarkable effectiveness compared to traditional feature-based learning-to-rank approaches, they suffer from severe efficiency limitations. 

Specifically, the self-attention mechanism exhibits quadratic time and memory complexity with respect to the input sequence length, whether over query-document pairs in pointwise settings or query and multi-documents in listwise settings, resulting in high computational cost and latency in practical deployments.

Token compression has emerged as a promising direction to improve the efficiency of these models~\cite{goyal2020powerbert,bolya2023token,wei2025deepseek}. 
Broadly speaking, these approaches aim at reducing the number of tokens (e.g., by aggregation) consumed by the models, leading to a latency improvement.
Recent work on dense retrieval models with bi-encoder architectures demonstrates that compressing token embeddings at the input layer can significantly reduce computational costs while preserving retrieval quality~\cite{zhang2025jasper}. 
However, the applicability of such compression strategies to cross-encoder rerankers,  
where query and document tokens interact directly and efficiency constraints are most severe, remains limited~\cite{Zhengmrlreranker} to pointwise ranking. 

In this paper, we study token compression applied to Transformer-based cross-encoder rerankers in pointwise and listwise settings. 
We start from the Jasper embedding-level compression approach~\cite{zhang2025jasper}, which applies 1D adaptive pooling to token embeddings in order to reduce the sequence length before the transformer layers.
Although being effective for bi-encoder embedding models, we find that the same strategy does not generalize well when applied to cross-encoder rerankers, and it leads to a degradation in performance. 
We hypothesize that the issue is due to the fundamental difference in interaction modeling. In cross-encoders, relevance signals emerge from fine-grained query/document token interactions that occur from the earliest transformer layers onward. Applying token compression at the embedding layer removes these signals before interaction can take place, leading to substantial performance degradation. Bi-encoders, by contrast, encode queries and documents independently and rely on vector-level similarity rather than explicit token interactions, making them less sensitive to early-stage token compression.

To address this limitation, we propose \textbf{Layer-wise Token Compression (LTC)}, which performs token pooling at an intermediate transformer layer. This design preserves early query/document token interactions in the lower layers, while reducing sequence length to accelerate the remaining computation. 
LTC yields a parametrizable efficiency/effectiveness trade-off: compressing earlier provides larger speedups but can harm ranking quality, whereas compressing later is safer but less beneficial. We systematically study this trade-off for pointwise cross-encoder rerankers and extend LTC to FIRST~\cite{gangi-reddy-etal-2024-first} style LLM-based listwise rerankers, where the much longer context can benefit even more from token compression.

Overall, we find that intermediate-layer compression preserves ranking quality while substantially reducing computation, naturally extends to listwise rerankers via selective document-token compression, and yields models that generalize better to long-document inputs.

\section{Related Work}
Improving the efficiency of Transformer-based models is a widely explored topic in the literature. Popular approaches include knowledge distillation~\cite{sanh2020distilbertdistilledversionbert,hinton2015distillingknowledgeneuralnetwork}, where a smaller model is trained to mimic a larger one; quantization~\cite{Zafrir_2019}, which reduces numerical precision to lower computation cost; and pruning~\cite{NEURIPS2019_2c601ad9}, which removes redundant parameters or attention heads.

\paragraph{Cross-encoder Reranking Models.} Cross-encoder rerankers score query-document pairs by jointly encoding the concatenated input, yielding strong effectiveness at the cost of higher latency~\cite{nogueira2019passage, bert2021yates,gao2021rethink,xu2026rankmambabenchmarkingmambasdocument,zhuang2021fastpassagererankingcontextualized}. Beyond pointwise scoring, recent work explores LLM-based listwise reranking, where models produce a ranking over a candidate set directly~\cite{sun2023chatgpt,ma2023zero,pradeep2023rankllm,gangi-reddy-etal-2024-first,chen2025first,setwise,chen2025attention,chen2026relevanceemergeslayerwisestudy}. These methods further increase input length (instruction + query + multiple documents), making efficiency a primary bottleneck.

\paragraph{Token Compression.} Recent methods accelerate transformers by reducing input sequence length, including token pruning based on attention statistics~\cite{goyal2020powerbert} and token merging that combines similar tokens~\cite{bolya2023token}. 
For dense document retrieval, recent work demonstrates that embedding-level token compression can be highly effective for bi-encoder retrievers by compressing token embeddings before deep transformer computation~\cite{zhang2025jasper}.
However, because cross-encoders rely on early query/document interactions, directly applying the same embedding-level compression can be detrimental. 
Token reduction has also been studied in the context of reranking models~\cite{Zhengmrlreranker}. Our work differs in two ways: (1) we systematically characterize where compression should occur in cross-encoder rerankers, showing that intermediate-layer compression offers a favorable efficiency/effectiveness trade-off; and (2) we extend this approach to listwise LLM rerankers. 

\paragraph{Connection to Other Compression Techniques}
Compression as a general technique has been broadly adopted by the applied ML community to lower computation cost~\cite{lecun1989optimalbraindamage,dettmers2022gpt3int8,frantar2023optq,xu-etal-2024-beyond-perplexity,li-etal-2023-compressing,jiang-etal-2023-llmlingua,chevalier-etal-2023-adapting,li2024promptcompressionlargelanguage,mu2023learningtocompresspromptswithgisttokens,ainslie-etal-2023-gqa}.
Model compression aims to reduce the model size for lower memory overhead and inference cost. Notable methods include model pruning~\cite{lecun1989optimalbraindamage,xia2024shearedllama}, model quantization~\cite{dettmers2022gpt3int8,frantar2023optq}, or distillation~\cite{hinton2015distillingknowledgeneuralnetwork,sanh2020distilbertdistilledversionbert}.
Similarly, prompt compression focuses on reducing the input length, thereby improving the efficiency~\cite{li2024promptcompressionlargelanguage}.
Due to the autoregressive nature of LLM inference, prompt compression can achieve notable speedup, especially with long texts.
KV cache compression takes a different approach. It aims to reduce the memory usage of KV cache during the autoregressive inference, by evicting/pruning less important vectors or storing them in lower numerical precision (i.e., KV cache quantization)~\cite{xiao2024efficientstreaming,cai2025pyramidkvdynamickvcache,liu2024kiviatuningfree,hooper2024kvquant}. Although sharing similar intuitions with prior prompt compression works, LTC focuses on the problem of improving reranker inference throughput, which is fundamentally different from LLM autoregressive inference. Moreover, LTC leads to orthogonal improvements, and it can be applied together with model or prompt compression.

\section{Layer-wise Token Compression}

Let $q$ be a query from a given distribution $\mathcal{Q}$ and let $\{d_i\}_{i=1}^k = \mathcal{D}_q \subset \mathcal{D}$ be a set of associated documents (e.g., retrieved from a retrieval engine) from a corpus $\mathcal{D}$.
The pointwise scoring function $s:\mathcal{Q}\times\mathcal{D} \rightarrow \mathbb{R}$ takes a query/document pair as input and returns a representation of likelihood of the document to be relevant for the input query, for any arbitrary definition of relevancy.
The pointwise (re)ranking task aims at finding the document $d^*$ with highest score, that is:
$d^* = \arg\max_{d_i\in\mathcal{D}_q} s(q, d_i)$.

In this work, the pointwise scoring function $s$ is implemented as a Transformer model that takes the query/document pair as input, concatenates it in a single sequence, 
and returns a relevancy score.

As a popular extension of pointwise ranking, the listwise scoring function provides the likelihood relevance of a document while reading all the others from the retrieved set $s:(q, d_i, \mathcal{D}_q)$.
In this work, we implement listwise ranking as a generative LLM that concatenates the query and all $k$ associated documents as input 
and generates ids of the documents ordered by likelihood relevance.

Let $\mathbf{H}^{(l)} \in \mathbb{R}^{n \times h}$ denote the hidden states at layer $l$,
where $n$ is the sequence length and $h$ is the hidden dimension. A compression module
$\mathcal{C}$ reduces the sequence length from $n$ to $n'=\lfloor n\cdot r\rfloor$, where $r \in (0,1]$ is the compression ratio. 

\begin{equation}
\mathbf{H}^{(l)}_{\text{compressed}} = \mathcal{C}(\mathbf{H}^{(l)}, r) \in \mathbb{R}^{n' \times h}.
\end{equation}

We implement $\mathcal{C}$ using 1D adaptive average pooling along the token dimension.
This operation merges neighboring token representations while preserving coarse semantic
structure. After compression, the attention mask and positional indices are updated for the shortened sequence.

We apply compression at a target layer $l^* \in \{1,\ldots,L\}$, where $L$ is the number of layers in the Transformer model.
Layers $1$ to $l^*-1$ process the full (uncompressed) sequence, while layers $l^*$ to $L$
operate on the compressed sequence. This delays pooling until after early layers capture
fine-grained query--document interactions, while reducing the quadratic attention cost in
later layers.

For listwise LLM rerankers, the input contains an instruction prompt, the query, and multiple
documents. To preserve the prompt and query structure, we apply compression only to document-token positions, and we introduce a \textbf{document mask}
$\mathbf{M}\in\{0,1\}^{n\times k}$.
The entry $(i,j)$ of the mask is 1 if the token $i$ belongs to the document $j$, 0 otherwise.
The mask prevents cross-document token compression, allowing the 1D pooling layer to use intra-document tokens only.

\begin{figure}[t]
\centering
\includegraphics[width=\columnwidth]{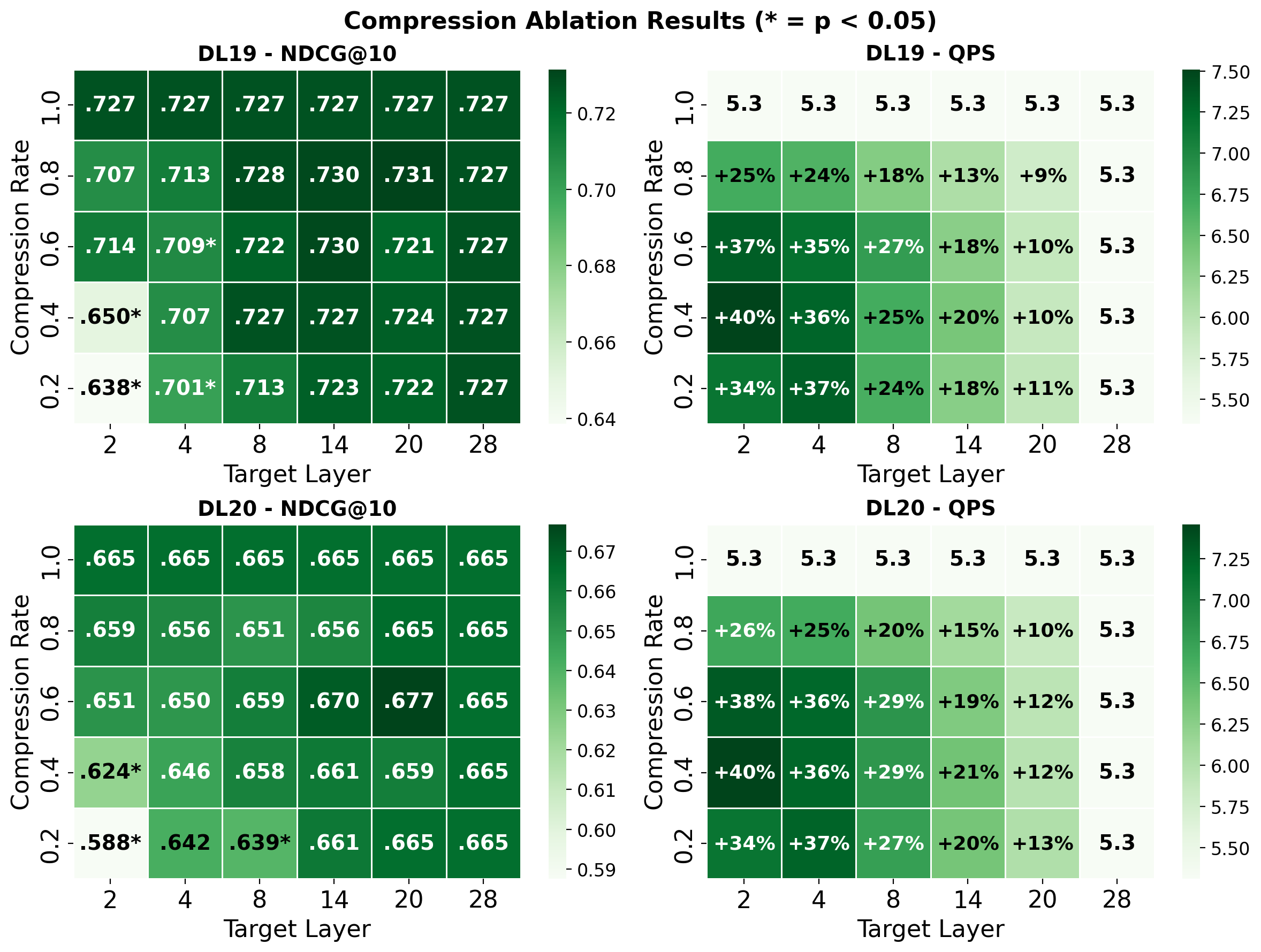}\vspace{-5pt}
\caption{Effect of LTC to pointwise passage ranking. Left: nDCG@10; Right: QPS. 
Compression rate 1.0 or Target Layer 28 means no compression applied. Cells with * denote statistically significant difference from no compression ($p < 0.05$).}
\Description{Effect of LTC to pointwise passage ranking. Left: nDCG@10; Right: QPS. 
Compression rate 1.0 or Target Layer 28 means no compression applied. Cells with * denote statistically significant difference from no compression ($p < 0.05$).}
\label{fig:pointwise_passage}
\end{figure}

\section{Experimental Setup}

\paragraph{Datasets.} We train all rerankers on the MS MARCO passage ranking dataset~\cite{bajaj2018msmarcohumangenerated}, which contains approximately 8.8 million passages and 500K training queries with sparse relevance judgments. For pointwise reranker training, we use training data provided by the Tevatron IR toolkit~\cite{tevatronv1,tevatronv2}, which contains hard negatives mined from strong dense and sparse retrievers.\footnote{\url{https://huggingface.co/datasets/Tevatron/msmarco-passage}}
 For listwise reranker training, we use the training data released with RankZephyr~\cite{RankZephyr} and FIRST~\cite{gangi-reddy-etal-2024-first}, which contains 20 documents ranked by GPT-4 as the gold ranking for each training query.\footnote{\url{https://huggingface.co/datasets/rryisthebest/rank_zephyr_training_data_alpha}}
 For evaluation, we use the TREC Deep Learning 2019 (DL19)~\cite{craswell2020overviewtrec2019deep} and 2020 (DL20)~\cite{craswell2021overviewtrec2020deep} test sets, which provide dense relevance annotations for both passage and document ranking tasks. To obtain candidate documents for reranking, we use BM25 implemented in Pyserini~\cite{pyserini} as the first-stage retriever and rerank the top-100 results. For passage ranking evaluation datasets, we truncate each passage to 128 tokens; for document ranking evaluation datasets, we truncate documents to 512 tokens to evaluate generalization to longer sequences. 

\paragraph{Models.} We experiment with two reranker architectures. For \emph{pointwise reranking}, we use Qwen3-0.6B-Base~\cite{qwen3},\footnote{\url{https://huggingface.co/Qwen/Qwen3-0.6B-Base}}
 a 28-layer decoder-only transformer, and add a linear classification head on top of the final hidden state to predict relevance scores. For \emph{listwise reranking}, we use Mistral-7B-Instruct-v0.3~\cite{jiang2023mistral},\footnote{\url{https://huggingface.co/mistralai/Mistral-7B-Instruct-v0.3}}
 a 32-layer model. In the listwise setting, the model receives $k=20$ candidate documents per query and uses the logits of the first generated token to produce document identifiers in relevance order. We then use a sliding window with a step size of 10, starting from the bottom of the ranking, to produce a full ranking over 100 documents.

\paragraph{Training.} For pointwise rerankers, we follow~\citet{gao2021rethink} and optimize cross-entropy loss over positive and negative documents, sampling one positive and five hard negatives per query. We train for 3 epochs with a batch size of 32 and a learning rate of $1\times10^{-5}$ using AdamW. For listwise rerankers, we combine language modeling loss with ListMLE ranking loss following FIRST~\cite{gangi-reddy-etal-2024-first,chen2025first}. 
In the listwise setting, we apply LoRA~\cite{hu2022lora} with rank 16 to adapt the model for ranking. 
All models are trained with compression applied during training to allow them to adapt to compressed representations.

\paragraph{Ablation Variables.} To characterize the efficiency--effectiveness trade-off, we systematically vary two hyperparameters: the compression rate $r \in \{0.2, 0.4, 0.6, 0.8, 1.0\}$, which controls the fraction of tokens retained after pooling, and the target layer $l^*$ at which compression is applied. For Qwen3-0.6B, we evaluate $l^* \in \{2, 4, 8, 14, 20\}$ out of 28 layers; for Mistral-7B, we evaluate $l^* \in \{4, 8\}$ out of 32 layers. A compression rate of $r=1.0$ corresponds to no compression (baseline).

\paragraph{Metrics.} We report nDCG@10 as the effectiveness metric, following standard TREC evaluation practice. For efficiency, we measure queries per second (QPS) on a single NVIDIA L40S GPU.

\subsection{Pointwise Reranking results}

\begin{figure}[t]
\centering
\includegraphics[width=\columnwidth]{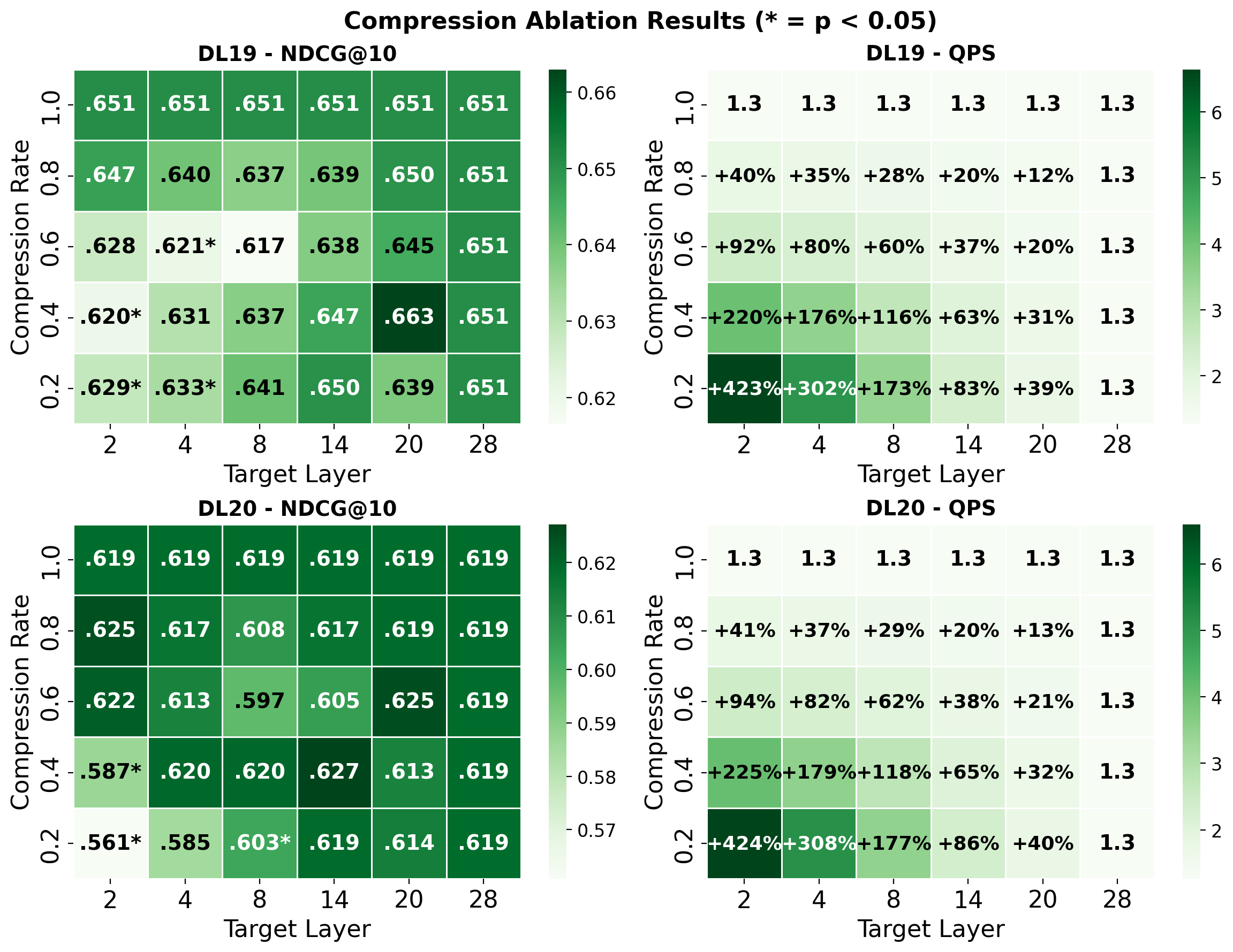}\vspace{-5pt}
\caption{Effect of LTC to pointwise document ranking. Left: nDCG@10; Right: QPS. 
Compression rate 1.0 or Target Layer 28 means no compression applied. Cells with * denote statistically significant difference from no compression ($p < 0.05$).}
\Description{Effect of LTC to pointwise document ranking. Left: nDCG@10; Right: QPS. 
Compression rate 1.0 or Target Layer 28 means no compression applied. Cells with * denote statistically significant difference from no compression ($p < 0.05$).}
\label{fig:pointwise_document}
\end{figure}

Figure~\ref{fig:pointwise_passage} presents the effectiveness/efficiency trade-off of LTC for Qwen3-0.6B on the passage ranking task across different compression rates and target layers. The heatmaps show nDCG@10 (left) and throughput in queries per second (right) for DL19 (first row) and DL20 (second row) test sets. Cells marked with * indicate statistically significant differences from the baseline ($p < 0.05$, paired $t$-test). The baseline without compression ($r=1.0$) achieves nDCG@10 of 0.727 on DL19 and 0.665 on DL20.

Several patterns emerge from these results. First, early-layer compression substantially degrades ranking quality. 
When applied at layer 2 with an aggressive compression rate ($r = 0.2$), nDCG@10 drops significantly to 0.638 ($p < 0.001$) on DL19, representing a 12.2\% degradation. 
Second, 
the nDCG@10 grows with both\crdel{,} compression rate and target layer, highlighting the importance of token-interaction signals from the lower layers. 
Middle-layer compression may achieve a favorable trade-off. 
When applied to layers 8-14, ranking quality remains stable even under substantial compression, and none of these configurations show statistically significant degradation. 
With $r=0.4$ at layer 8, the model retains nDCG@10 of 0.7274 on DL19, virtually identical to the baseline, while increasing throughput from 5.35 to 6.69 QPS (25\% improvement).
Third, light compression at later layers can slightly exceed baseline performance. At layer 14-20 with $r=0.8$, nDCG@10 reaches 0.7314 on DL19, marginally surpassing the uncompressed model (but not statistically significant). 
We hypothesize that moderate compression acts as a regularizer, smoothing over noisy token-level variations.

\paragraph{Jasper compression}
We compared our approach against Jasper encoder~\cite{zhang2025jasper}, which performs compression at the embedding layer with an adaptive compression rate of 0.5. 
On DL19 and DL20 passage ranking test sets, Jasper compression shows QPS values of 6.52 and 6.51, representing a 25\% relative improvement, and nDCG@10 of 0.5832 and 0.4975 respectively. 
At same QPS gain, our approach shows 0.727 (DL19, $r=0.4$, $l=8$, QPS +25\%) and 0.651 (DL20, $r=0.8$, $l=4$, QPS +25\%) nDCG@10.
This result confirms our hypothesis that early transformer layers capture fine-grained query/document matching signals that are destroyed by premature token pooling. 

\paragraph{Generalization to Document Ranking.} A key question is whether rerankers trained on short passages generalize to longer documents. Figure~\ref{fig:pointwise_document} shows results when applying passage-trained models to the document ranking task with 512-token inputs.

Notably, models trained with compression achieve competitive or even superior effectiveness compared to the uncompressed baseline, despite never seeing long documents during training. On DL19, the compressed model with $r=0.4$ at layer 20 achieves nDCG@10 of 0.663, compared to 0.651 for the baseline. This is consistent with LTC acting as a regularizer that encourages more robust, length-invariant representations rather than overfitting to passage-specific patterns. We treat this interpretation as a hypothesis: a deeper representation-level analysis (e.g., probing hidden-state norms or attention sparsity across input lengths) is left to future work.

\begin{figure}[t]
\centering
\includegraphics[width=.75\columnwidth]{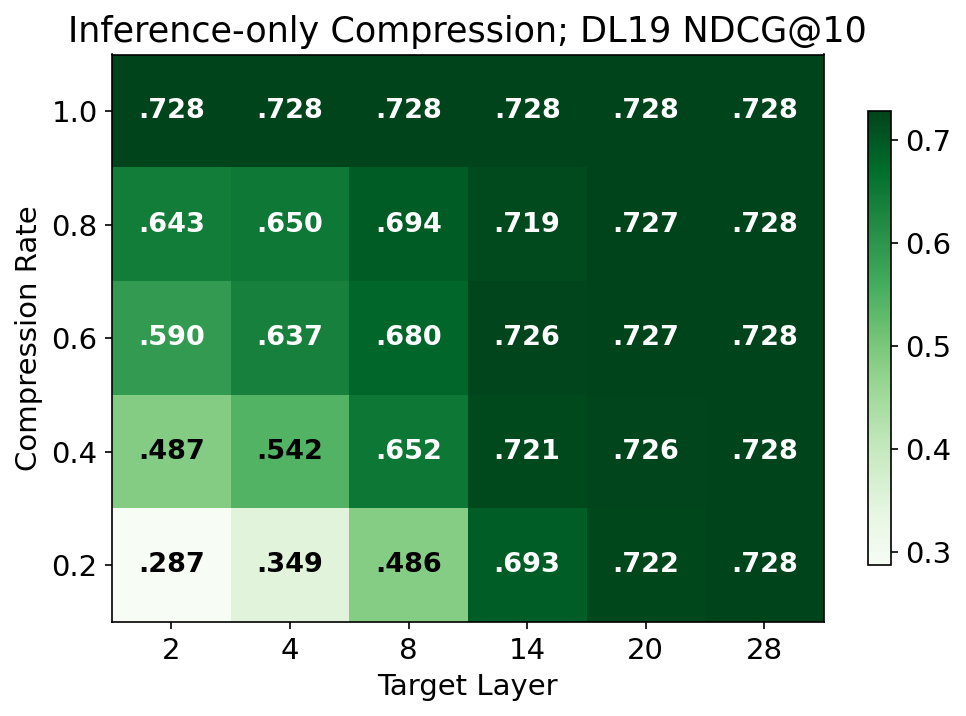}
\vspace{-10pt}\caption{Effect of zero-shot compression for pointwise passage ranking. LTC is applied to a model trained on MS~MARCO without compression. Compression rate 1.0 or Target Layer 28 means no compression applied.}
\Description{Effect of zero-shot compression for pointwise passage ranking. LTC is applied to a model trained on MS~MARCO without compression. Compression rate 1.0 or Target Layer 28 means no compression applied.}
\label{fig:inference-only}\vspace{-5pt}
\end{figure}

\paragraph{Pointwise zero-shot}
A natural question is whether the performance gains of LTC require compression-aware training, or whether similar efficiency benefits can be obtained by simply applying token compression at inference time to a model trained without compression. To investigate this, we re-trained the uncompressed baseline model and apply LTC with varying compression rates and target layers at inference time only, without any additional fine-tuning.
Figure~\ref{fig:inference-only} presents the results of this inference-only compression ablation on DL19 passage ranking. The contrast with the trained LTC results (Figure~\ref{fig:pointwise_passage}) is striking. While trained LTC maintains near-baseline effectiveness even under aggressive compression (e.g., $r$=0.4, $\ell$=8 achieves 0.727 nDCG@10 compared to the 0.727 baseline), inference-only compression degrades rapidly as compression becomes more aggressive. At $r$=0.2, $\ell$=2, inference-only compression drops to 0.287 nDCG@10: a catastrophic 60\% relative decrease from the 0.728 baseline when compared to 0.638 achieved by the same model fine-tuned with LTC.
The degradation follows a clear pattern: performance is preserved when compression starts at later layers ($\ell \geq 14$) regardless of rate, but collapses when applied at early layers with aggressive rates. For example, at $r$=0.8, $\ell$=20, inference-only compression achieves 0.727, nearly matching the baseline. However, at the same rate with $\ell$=2, performance drops to 0.643. This gap widens dramatically at lower rates: $r$=0.2 at $\ell$=20 still achieves 0.722, but at $\ell$=4 it falls to 0.349.
These results demonstrate that compression-aware training is essential for LTC to be effective. The trained models learn to produce representations that are robust to token reduction at the designated layers, whereas the uncompressed model's intermediate representations are not structured to tolerate such information loss. This finding underscores that LTC is not merely a post-hoc pruning technique, but a learned compression strategy where the model adapts its internal representations to accommodate reduced token sequences.

\begin{table}[t]
\centering
\caption{Zero-shot BEIR evaluation (nDCG@10). Models trained on MS~MARCO (passages) with different LTC compression configurations. BM25 top-100 candidates reranked.}
\label{tab:beir}
\resizebox{\columnwidth}{!}{%
\setlength{\tabcolsep}{5pt}
\begin{tabular}{lccccc}
\toprule
Dataset & BM25 & No Comp. & $r$=0.8, $\ell$=20 & $r$=0.6, $\ell$=14 & $r$=0.4, $\ell$=8 \\
\midrule
NFCorpus & .322 & \textbf{.342} & .338 & .338 & .331 \\
FiQA & .236 & .323 & \textbf{.332} & .322 & .317 \\
ScIDocs & .149 & .170 & \textbf{.173} & .170 & .164 \\
SciFact & .679 & \textbf{.685} & .675 & .668 & .666 \\
TREC-COVID & .595 & .734 & \textbf{.749} & .738 & .739 \\
ArguAna & \textbf{.408} & .280 & .278 & .282 & .278 \\
\midrule
Avg. & .398 & .422 & \textbf{.424} & .420 & .416 \\
QPS. & - & +0\% & +9\% & +18\% & +25\% \\
\bottomrule
\end{tabular}}
\end{table}

\paragraph{BEIR evaluation}
To evaluate whether LTC-compressed models generalize beyond the MS~MARCO training domain, we conduct zero-shot evaluation on six diverse BEIR datasets: NFCorpus, FiQA, ScIDocs, SciFact, TREC-COVID, and ArguAna. For each dataset, we retrieve the top-100 candidates using BM25 and rerank them with our trained models. No dataset-specific fine-tuning is performed.
Table~\ref{tab:beir} presents the results. Across all compressed configurations, zero-shot performance remains remarkably close to the uncompressed baseline. The mildly compressed model ($r$=0.8, $\ell$=20) achieves the highest average nDCG@10 of 0.424, slightly exceeding the uncompressed baseline (0.422). It outperforms the baseline on three of six datasets (FiQA, ScIDocs, and TREC-COVID), suggesting that mild compression may act as a form of regularization that benefits out-of-domain generalization.
Even the most aggressively compressed model ($r$=0.4, $\ell$=8) maintains competitive performance with an average of 0.416, representing only a 1.4\% relative decrease from the baseline despite retaining only 40\% of tokens from layer 8 onward. The moderate configuration ($r$=0.6, $\ell$=14) achieves 0.420, closely tracking the baseline across all datasets.
All reranking configurations substantially outperform the BM25 first-stage retrieval (0.398 average), confirming that the compressed models retain their reranking capability in zero-shot settings. The one exception is ArguAna, where all rerankers underperform BM25. However, this is a known challenge for pointwise rerankers on this counter-argument retrieval task, where relevant documents share vocabulary but differ in stance.

\subsection{Listwise Reranking Results}

\begin{figure}[t]
\centering
\includegraphics[width=\columnwidth]{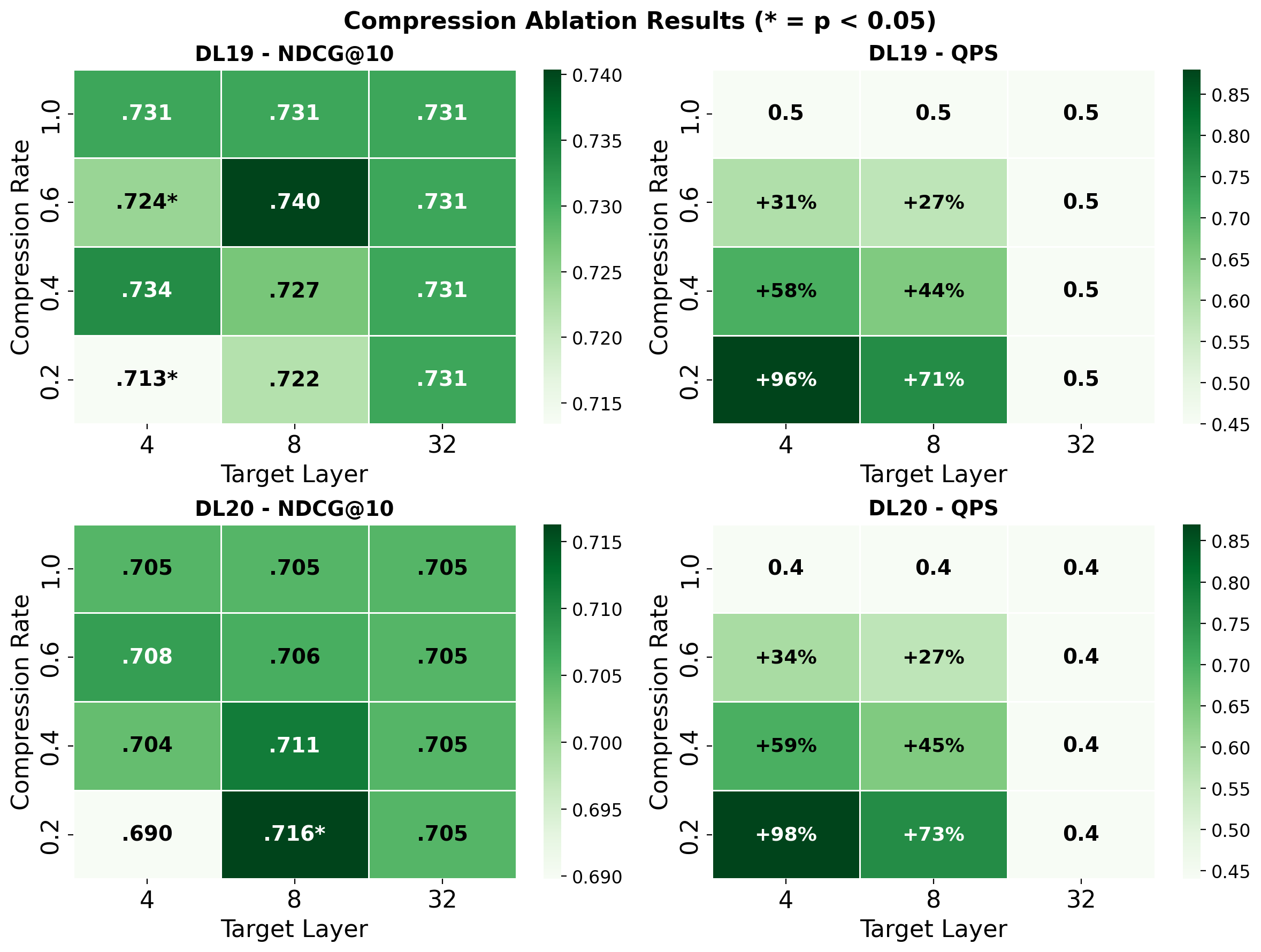}\vspace{-5pt}
\caption{Effect of LTC to listwise passage ranking. Left: nDCG@10; Right: QPS. Cells with * denote statistically significant difference from no compression ($p < 0.05$).}
\Description{Effect of LTC to listwise passage ranking. Left: nDCG@10; Right: QPS. Cells with * denote statistically significant difference from no compression ($p < 0.05$).}
\label{fig:listwise_passage}
\end{figure}

\begin{figure}[t]
\centering
\includegraphics[width=\columnwidth]{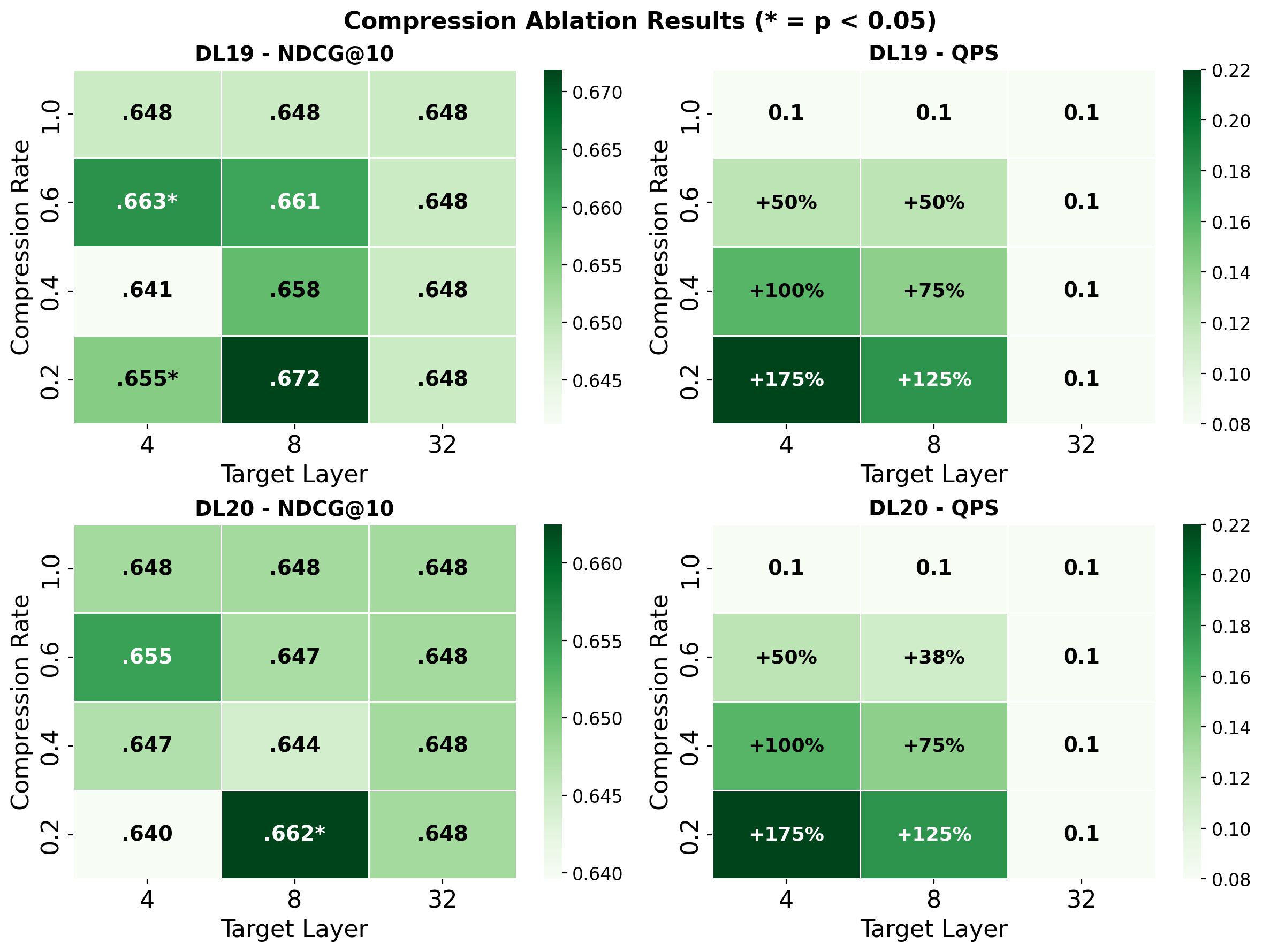}\vspace{-5pt}
\caption{Effect of LTC to listwise document ranking. Left: nDCG@10; Right: QPS. Cells with * denote statistically significant difference from no compression ($p < 0.05$).}
\Description{Effect of LTC to listwise document ranking. Left: nDCG@10; Right: QPS. Cells with * denote statistically significant difference from no compression ($p < 0.05$).}
\label{fig:listwise_document}\vspace{-5pt}
\end{figure}

Figure~\ref{fig:listwise_passage} and~\ref{fig:listwise_document} present results for the Mistral-7B listwise reranker on passage and document ranking respectively.
As described before, when applied to listwise, LTC doesn't apply cross-document compression and it preserves query and instruction tokens.

Regarding the passage ranking tasks, most of the listwise LTC configurations improve effectiveness over the non-compression baseline, with some settings achieving statistically significant gains $p < 0.05$. 
It is also worth noting that, for all configurations that do not incur a statistically significant decrease in nDCG@10, LTC is associated with higher throughput, with gains ranging from +27\% to +73\%.

\section{Conclusion}

We presented Layer-wise Token Compression (LTC), a simple yet effective approach for efficient transformer-based document reranking models. Our key insight is that token compression should be applied at intermediate transformer layers rather than at the input embedding level, allowing early layers to capture fine-grained query/document interactions before compressing representations for efficient downstream processing. 
Our experiments yield three main findings. First, compression at middle layers (8-14) preserves ranking effectiveness (no statistically significant degradation) while achieving 25-116\% throughput improvement. 
Second, early-layer compression (layer 2-4) significantly harms ranking quality, suggesting that initial layers encode critical matching signals. 
Third, and perhaps most surprisingly, models trained with LTC generalize better to longer document sequences than uncompressed models, often achieving superior effectiveness on out-of-distribution lengths. 
The method extends naturally to listwise LLM rerankers through selective compression, where only document tokens are compressed while preserving query and instruction tokens. This enables nearly 2$\times$ speedup on 7B-parameter models with minimal effectiveness loss.

\paragraph{Limitations and Future Work.}
Our contribution is methodological: we identify \emph{where} in the architecture compression should be applied, while the underlying operator is intentionally a simple adaptive average pool. Several limitations follow.
(i) The target layer $l^*$ and rate $r$ are selected by manual sweeps; automating this choice (e.g., via gradient-based search or neural architecture search) would ease deployment on new model families.
(ii) Our empirical comparison focuses on Jasper-style embedding compression~\cite{zhang2025jasper}; broader empirical comparisons with peer-reviewed alternatives such as MRL-Reranker~\cite{Zhengmrlreranker}, prompt compression~\cite{li2024promptcompressionlargelanguage}, KV-cache compression~\cite{cai2025pyramidkvdynamickvcache,liu2024kiviatuningfree,hooper2024kvquant}, early-exit, and list-truncation strategies are left to future work; we expect LTC to be largely orthogonal to these techniques and combinable with them.
(iii) Although mild compression at later layers can be applied without retraining (Fig.~\ref{fig:inference-only}), aggressive configurations still benefit from compression-aware fine-tuning; fully training-free LTC variants for existing rerankers remain open. (iv) The regularization interpretation is a hypothesis motivated by our results; a representation-level analysis (e.g., hidden-state norms and attention sparsity as a function of input length) and a runtime breakdown of the pooling operator versus downstream-layer savings are valuable directions for future work.

\bibliographystyle{ACM-Reference-Format}
\balance
\bibliography{mybib}


\begin{thebibliography}{52}


\ifx \showCODEN    \undefined \def \showCODEN     #1{\unskip}     \fi
\ifx \showISBNx    \undefined \def \showISBNx     #1{\unskip}     \fi
\ifx \showISBNxiii \undefined \def \showISBNxiii  #1{\unskip}     \fi
\ifx \showISSN     \undefined \def \showISSN      #1{\unskip}     \fi
\ifx \showLCCN     \undefined \def \showLCCN      #1{\unskip}     \fi
\ifx \shownote     \undefined \def \shownote      #1{#1}          \fi
\ifx \showarticletitle \undefined \def \showarticletitle #1{#1}   \fi
\ifx \showURL      \undefined \def \showURL       {\relax}        \fi
\providecommand\bibfield[2]{#2}
\providecommand\bibinfo[2]{#2}
\providecommand\natexlab[1]{#1}
\providecommand\showeprint[2][]{arXiv:#2}

\bibitem[Ainslie et~al\mbox{.}(2023)]%
        {ainslie-etal-2023-gqa}
\bibfield{author}{\bibinfo{person}{Joshua Ainslie}, \bibinfo{person}{James Lee-Thorp}, \bibinfo{person}{Michiel de Jong}, \bibinfo{person}{Yury Zemlyanskiy}, \bibinfo{person}{Federico Lebron}, {and} \bibinfo{person}{Sumit Sanghai}.} \bibinfo{year}{2023}\natexlab{}.
\newblock \showarticletitle{{GQA}: Training Generalized Multi-Query Transformer Models from Multi-Head Checkpoints}. In \bibinfo{booktitle}{\emph{Proceedings of the 2023 Conference on Empirical Methods in Natural Language Processing}}, \bibfield{editor}{\bibinfo{person}{Houda Bouamor}, \bibinfo{person}{Juan Pino}, {and} \bibinfo{person}{Kalika Bali}} (Eds.). \bibinfo{publisher}{Association for Computational Linguistics}, \bibinfo{address}{Singapore}, \bibinfo{pages}{4895--4901}.
\newblock
\href{https://doi.org/10.18653/v1/2023.emnlp-main.298}{doi:\nolinkurl{10.18653/v1/2023.emnlp-main.298}}


\bibitem[Bajaj et~al\mbox{.}(2018)]%
        {bajaj2018msmarcohumangenerated}
\bibfield{author}{\bibinfo{person}{Payal Bajaj}, \bibinfo{person}{Daniel Campos}, \bibinfo{person}{Nick Craswell}, \bibinfo{person}{Li Deng}, \bibinfo{person}{Jianfeng Gao}, \bibinfo{person}{Xiaodong Liu}, \bibinfo{person}{Rangan Majumder}, \bibinfo{person}{Andrew McNamara}, \bibinfo{person}{Bhaskar Mitra}, \bibinfo{person}{Tri Nguyen}, \bibinfo{person}{Mir Rosenberg}, \bibinfo{person}{Xia Song}, \bibinfo{person}{Alina Stoica}, \bibinfo{person}{Saurabh Tiwary}, {and} \bibinfo{person}{Tong Wang}.} \bibinfo{year}{2018}\natexlab{}.
\newblock \bibinfo{title}{MS MARCO: A Human Generated MAchine Reading COmprehension Dataset}.
\newblock
\showeprint[arxiv]{1611.09268}~[cs.CL]
\urldef\tempurl%
\url{https://arxiv.org/abs/1611.09268}
\showURL{%
\tempurl}


\bibitem[Bolya et~al\mbox{.}(2023)]%
        {bolya2023token}
\bibfield{author}{\bibinfo{person}{Daniel Bolya}, \bibinfo{person}{Cheng-Yang Fu}, \bibinfo{person}{Xiaoliang Dai}, \bibinfo{person}{Peizhao Zhang}, \bibinfo{person}{Christoph Feichtenhofer}, {and} \bibinfo{person}{Judy Hoffman}.} \bibinfo{year}{2023}\natexlab{}.
\newblock \showarticletitle{Token Merging: Your ViT But Faster}. In \bibinfo{booktitle}{\emph{The Eleventh International Conference on Learning Representations}}.
\newblock
\urldef\tempurl%
\url{https://openreview.net/forum?id=JroZRaRw7Eu}
\showURL{%
\tempurl}


\bibitem[Cai et~al\mbox{.}(2025)]%
        {cai2025pyramidkvdynamickvcache}
\bibfield{author}{\bibinfo{person}{Zefan Cai}, \bibinfo{person}{Yichi Zhang}, \bibinfo{person}{Bofei Gao}, \bibinfo{person}{Yuliang Liu}, \bibinfo{person}{Yucheng Li}, \bibinfo{person}{Tianyu Liu}, \bibinfo{person}{Keming Lu}, \bibinfo{person}{Wayne Xiong}, \bibinfo{person}{Yue Dong}, \bibinfo{person}{Junjie Hu}, {and} \bibinfo{person}{Wen Xiao}.} \bibinfo{year}{2025}\natexlab{}.
\newblock \bibinfo{title}{PyramidKV: Dynamic KV Cache Compression based on Pyramidal Information Funneling}.
\newblock
\showeprint[arxiv]{2406.02069}~[cs.CL]
\urldef\tempurl%
\url{https://arxiv.org/abs/2406.02069}
\showURL{%
\tempurl}


\bibitem[Chen et~al\mbox{.}(2026)]%
        {chen2026relevanceemergeslayerwisestudy}
\bibfield{author}{\bibinfo{person}{Haodong Chen}, \bibinfo{person}{Shengyao Zhuang}, \bibinfo{person}{Zheng Yao}, \bibinfo{person}{Guido Zuccon}, {and} \bibinfo{person}{Teerapong Leelanupab}.} \bibinfo{year}{2026}\natexlab{}.
\newblock \bibinfo{title}{Where Relevance Emerges: A Layer-Wise Study of Internal Attention for Zero-Shot Re-Ranking}.
\newblock
\showeprint[arxiv]{2602.22591}~[cs.IR]
\urldef\tempurl%
\url{https://arxiv.org/abs/2602.22591}
\showURL{%
\tempurl}


\bibitem[Chen et~al\mbox{.}(2025a)]%
        {chen2025attention}
\bibfield{author}{\bibinfo{person}{Shijie Chen}, \bibinfo{person}{Bernal~Jimenez Gutierrez}, {and} \bibinfo{person}{Yu Su}.} \bibinfo{year}{2025}\natexlab{a}.
\newblock \showarticletitle{Attention in Large Language Models Yields Efficient Zero-Shot Re-Rankers}. In \bibinfo{booktitle}{\emph{The Thirteenth International Conference on Learning Representations}}.
\newblock
\urldef\tempurl%
\url{https://openreview.net/forum?id=yzloNYH3QN}
\showURL{%
\tempurl}


\bibitem[Chen et~al\mbox{.}(2025b)]%
        {chen2025first}
\bibfield{author}{\bibinfo{person}{Zijian Chen}, \bibinfo{person}{Ronak Pradeep}, {and} \bibinfo{person}{Jimmy Lin}.} \bibinfo{year}{2025}\natexlab{b}.
\newblock \showarticletitle{Accelerating Listwise Reranking: Reproducing and Enhancing FIRST}. In \bibinfo{booktitle}{\emph{Proceedings of the 48th International ACM SIGIR Conference on Research and Development in Information Retrieval}} (Padua, Italy) \emph{(\bibinfo{series}{SIGIR '25})}. \bibinfo{publisher}{Association for Computing Machinery}, \bibinfo{address}{New York, NY, USA}, \bibinfo{pages}{3165–3172}.
\newblock
\showISBNx{9798400715921}
\href{https://doi.org/10.1145/3726302.3730287}{doi:\nolinkurl{10.1145/3726302.3730287}}


\bibitem[Chevalier et~al\mbox{.}(2023)]%
        {chevalier-etal-2023-adapting}
\bibfield{author}{\bibinfo{person}{Alexis Chevalier}, \bibinfo{person}{Alexander Wettig}, \bibinfo{person}{Anirudh Ajith}, {and} \bibinfo{person}{Danqi Chen}.} \bibinfo{year}{2023}\natexlab{}.
\newblock \showarticletitle{Adapting Language Models to Compress Contexts}. In \bibinfo{booktitle}{\emph{Proceedings of the 2023 Conference on Empirical Methods in Natural Language Processing}}, \bibfield{editor}{\bibinfo{person}{Houda Bouamor}, \bibinfo{person}{Juan Pino}, {and} \bibinfo{person}{Kalika Bali}} (Eds.). \bibinfo{publisher}{Association for Computational Linguistics}, \bibinfo{address}{Singapore}, \bibinfo{pages}{3829--3846}.
\newblock
\href{https://doi.org/10.18653/v1/2023.emnlp-main.232}{doi:\nolinkurl{10.18653/v1/2023.emnlp-main.232}}


\bibitem[Craswell et~al\mbox{.}(2021)]%
        {craswell2021overviewtrec2020deep}
\bibfield{author}{\bibinfo{person}{Nick Craswell}, \bibinfo{person}{Bhaskar Mitra}, \bibinfo{person}{Emine Yilmaz}, {and} \bibinfo{person}{Daniel Campos}.} \bibinfo{year}{2021}\natexlab{}.
\newblock \bibinfo{title}{Overview of the TREC 2020 deep learning track}.
\newblock
\showeprint[arxiv]{2102.07662}~[cs.IR]
\urldef\tempurl%
\url{https://arxiv.org/abs/2102.07662}
\showURL{%
\tempurl}


\bibitem[Craswell et~al\mbox{.}(2020)]%
        {craswell2020overviewtrec2019deep}
\bibfield{author}{\bibinfo{person}{Nick Craswell}, \bibinfo{person}{Bhaskar Mitra}, \bibinfo{person}{Emine Yilmaz}, \bibinfo{person}{Daniel Campos}, {and} \bibinfo{person}{Ellen~M. Voorhees}.} \bibinfo{year}{2020}\natexlab{}.
\newblock \bibinfo{title}{Overview of the TREC 2019 deep learning track}.
\newblock
\showeprint[arxiv]{2003.07820}~[cs.IR]
\urldef\tempurl%
\url{https://arxiv.org/abs/2003.07820}
\showURL{%
\tempurl}


\bibitem[Dettmers et~al\mbox{.}(2022)]%
        {dettmers2022gpt3int8}
\bibfield{author}{\bibinfo{person}{Tim Dettmers}, \bibinfo{person}{Mike Lewis}, \bibinfo{person}{Younes Belkada}, {and} \bibinfo{person}{Luke Zettlemoyer}.} \bibinfo{year}{2022}\natexlab{}.
\newblock \showarticletitle{Gpt3. int8 (): 8-bit matrix multiplication for transformers at scale}.
\newblock \bibinfo{journal}{\emph{Advances in neural information processing systems}}  \bibinfo{volume}{35} (\bibinfo{year}{2022}), \bibinfo{pages}{30318--30332}.
\newblock


\bibitem[Frantar et~al\mbox{.}(2023)]%
        {frantar2023optq}
\bibfield{author}{\bibinfo{person}{Elias Frantar}, \bibinfo{person}{Saleh Ashkboos}, \bibinfo{person}{Torsten Hoefler}, {and} \bibinfo{person}{Dan Alistarh}.} \bibinfo{year}{2023}\natexlab{}.
\newblock \showarticletitle{{OPTQ}: Accurate Quantization for Generative Pre-trained Transformers}. In \bibinfo{booktitle}{\emph{The Eleventh International Conference on Learning Representations}}.
\newblock
\urldef\tempurl%
\url{https://openreview.net/forum?id=tcbBPnfwxS}
\showURL{%
\tempurl}


\bibitem[Gangi~Reddy et~al\mbox{.}(2024)]%
        {gangi-reddy-etal-2024-first}
\bibfield{author}{\bibinfo{person}{Revanth Gangi~Reddy}, \bibinfo{person}{JaeHyeok Doo}, \bibinfo{person}{Yifei Xu}, \bibinfo{person}{Md~Arafat Sultan}, \bibinfo{person}{Deevya Swain}, \bibinfo{person}{Avirup Sil}, {and} \bibinfo{person}{Heng Ji}.} \bibinfo{year}{2024}\natexlab{}.
\newblock \showarticletitle{{FIRST}: Faster Improved Listwise Reranking with Single Token Decoding}. In \bibinfo{booktitle}{\emph{Proceedings of the 2024 Conference on Empirical Methods in Natural Language Processing}}, \bibfield{editor}{\bibinfo{person}{Yaser Al-Onaizan}, \bibinfo{person}{Mohit Bansal}, {and} \bibinfo{person}{Yun-Nung Chen}} (Eds.). \bibinfo{publisher}{Association for Computational Linguistics}, \bibinfo{address}{Miami, Florida, USA}, \bibinfo{pages}{8642--8652}.
\newblock
\href{https://doi.org/10.18653/v1/2024.emnlp-main.491}{doi:\nolinkurl{10.18653/v1/2024.emnlp-main.491}}


\bibitem[Gao et~al\mbox{.}(2021)]%
        {gao2021rethink}
\bibfield{author}{\bibinfo{person}{Luyu Gao}, \bibinfo{person}{Zhuyun Dai}, {and} \bibinfo{person}{Jamie Callan}.} \bibinfo{year}{2021}\natexlab{}.
\newblock \showarticletitle{Rethink Training of BERT Rerankers in Multi-stage Retrieval Pipeline}. In \bibinfo{booktitle}{\emph{Advances in Information Retrieval: 43rd European Conference on IR Research, ECIR 2021, Virtual Event, March 28 – April 1, 2021, Proceedings, Part II}}. \bibinfo{publisher}{Springer-Verlag}, \bibinfo{address}{Berlin, Heidelberg}, \bibinfo{pages}{280–286}.
\newblock
\showISBNx{978-3-030-72239-5}
\href{https://doi.org/10.1007/978-3-030-72240-1_26}{doi:\nolinkurl{10.1007/978-3-030-72240-1_26}}


\bibitem[Gao et~al\mbox{.}(2023)]%
        {tevatronv1}
\bibfield{author}{\bibinfo{person}{Luyu Gao}, \bibinfo{person}{Xueguang Ma}, \bibinfo{person}{Jimmy Lin}, {and} \bibinfo{person}{Jamie Callan}.} \bibinfo{year}{2023}\natexlab{}.
\newblock \showarticletitle{Tevatron: An Efficient and Flexible Toolkit for Neural Retrieval}. In \bibinfo{booktitle}{\emph{Proceedings of the 46th International ACM SIGIR Conference on Research and Development in Information Retrieval}} (Taipei, Taiwan) \emph{(\bibinfo{series}{SIGIR '23})}. \bibinfo{publisher}{Association for Computing Machinery}, \bibinfo{address}{New York, NY, USA}, \bibinfo{pages}{3120–3124}.
\newblock
\showISBNx{9781450394086}
\href{https://doi.org/10.1145/3539618.3591805}{doi:\nolinkurl{10.1145/3539618.3591805}}


\bibitem[Goyal et~al\mbox{.}(2020)]%
        {goyal2020powerbert}
\bibfield{author}{\bibinfo{person}{Saurabh Goyal}, \bibinfo{person}{Anamitra~Roy Choudhury}, \bibinfo{person}{Saurabh~M. Raje}, \bibinfo{person}{Venkatesan~T. Chakaravarthy}, \bibinfo{person}{Yogish Sabharwal}, {and} \bibinfo{person}{Ashish Verma}.} \bibinfo{year}{2020}\natexlab{}.
\newblock \showarticletitle{PoWER-BERT: accelerating BERT inference via progressive word-vector elimination}. In \bibinfo{booktitle}{\emph{Proceedings of the 37th International Conference on Machine Learning}} \emph{(\bibinfo{series}{ICML'20})}. \bibinfo{publisher}{JMLR.org}, Article \bibinfo{articleno}{346}, \bibinfo{numpages}{10}~pages.
\newblock


\bibitem[Hinton et~al\mbox{.}(2015)]%
        {hinton2015distillingknowledgeneuralnetwork}
\bibfield{author}{\bibinfo{person}{Geoffrey Hinton}, \bibinfo{person}{Oriol Vinyals}, {and} \bibinfo{person}{Jeff Dean}.} \bibinfo{year}{2015}\natexlab{}.
\newblock \bibinfo{title}{Distilling the Knowledge in a Neural Network}.
\newblock
\showeprint[arxiv]{1503.02531}~[stat.ML]
\urldef\tempurl%
\url{https://arxiv.org/abs/1503.02531}
\showURL{%
\tempurl}


\bibitem[Hooper et~al\mbox{.}(2024)]%
        {hooper2024kvquant}
\bibfield{author}{\bibinfo{person}{Coleman Hooper}, \bibinfo{person}{Sehoon Kim}, \bibinfo{person}{Hiva Mohammadzadeh}, \bibinfo{person}{Michael~W Mahoney}, \bibinfo{person}{Yakun~S Shao}, \bibinfo{person}{Kurt Keutzer}, {and} \bibinfo{person}{Amir Gholami}.} \bibinfo{year}{2024}\natexlab{}.
\newblock \showarticletitle{Kvquant: Towards 10 million context length llm inference with kv cache quantization}.
\newblock \bibinfo{journal}{\emph{Advances in Neural Information Processing Systems}}  \bibinfo{volume}{37} (\bibinfo{year}{2024}), \bibinfo{pages}{1270--1303}.
\newblock


\bibitem[Hu et~al\mbox{.}(2022)]%
        {hu2022lora}
\bibfield{author}{\bibinfo{person}{Edward~J Hu}, \bibinfo{person}{yelong shen}, \bibinfo{person}{Phillip Wallis}, \bibinfo{person}{Zeyuan Allen-Zhu}, \bibinfo{person}{Yuanzhi Li}, \bibinfo{person}{Shean Wang}, \bibinfo{person}{Lu Wang}, {and} \bibinfo{person}{Weizhu Chen}.} \bibinfo{year}{2022}\natexlab{}.
\newblock \showarticletitle{Lo{RA}: Low-Rank Adaptation of Large Language Models}. In \bibinfo{booktitle}{\emph{International Conference on Learning Representations}}.
\newblock
\urldef\tempurl%
\url{https://openreview.net/forum?id=nZeVKeeFYf9}
\showURL{%
\tempurl}


\bibitem[Jiang et~al\mbox{.}(2023a)]%
        {jiang2023mistral}
\bibfield{author}{\bibinfo{person}{Albert~Q. Jiang}, \bibinfo{person}{Alexandre Sablayrolles}, \bibinfo{person}{Arthur Mensch}, \bibinfo{person}{Chris Bamford}, \bibinfo{person}{Devendra~Singh Chaplot}, \bibinfo{person}{Diego de~las Casas}, \bibinfo{person}{Florian Bressand}, \bibinfo{person}{Gianna Lengyel}, \bibinfo{person}{Guillaume Lample}, \bibinfo{person}{Lucile Saulnier}, \bibinfo{person}{Lélio~Renard Lavaud}, \bibinfo{person}{Marie-Anne Lachaux}, \bibinfo{person}{Pierre Stock}, \bibinfo{person}{Teven~Le Scao}, \bibinfo{person}{Thibaut Lavril}, \bibinfo{person}{Thomas Wang}, \bibinfo{person}{Timothée Lacroix}, {and} \bibinfo{person}{William~El Sayed}.} \bibinfo{year}{2023}\natexlab{a}.
\newblock \bibinfo{title}{Mistral 7B}.
\newblock
\showeprint[arxiv]{2310.06825}~[cs.CL]
\urldef\tempurl%
\url{https://arxiv.org/abs/2310.06825}
\showURL{%
\tempurl}


\bibitem[Jiang et~al\mbox{.}(2023b)]%
        {jiang-etal-2023-llmlingua}
\bibfield{author}{\bibinfo{person}{Huiqiang Jiang}, \bibinfo{person}{Qianhui Wu}, \bibinfo{person}{Chin-Yew Lin}, \bibinfo{person}{Yuqing Yang}, {and} \bibinfo{person}{Lili Qiu}.} \bibinfo{year}{2023}\natexlab{b}.
\newblock \showarticletitle{{LLML}ingua: Compressing Prompts for Accelerated Inference of Large Language Models}. In \bibinfo{booktitle}{\emph{Proceedings of the 2023 Conference on Empirical Methods in Natural Language Processing}}, \bibfield{editor}{\bibinfo{person}{Houda Bouamor}, \bibinfo{person}{Juan Pino}, {and} \bibinfo{person}{Kalika Bali}} (Eds.). \bibinfo{publisher}{Association for Computational Linguistics}, \bibinfo{address}{Singapore}, \bibinfo{pages}{13358--13376}.
\newblock
\href{https://doi.org/10.18653/v1/2023.emnlp-main.825}{doi:\nolinkurl{10.18653/v1/2023.emnlp-main.825}}


\bibitem[LeCun et~al\mbox{.}(1989)]%
        {lecun1989optimalbraindamage}
\bibfield{author}{\bibinfo{person}{Yann LeCun}, \bibinfo{person}{John Denker}, {and} \bibinfo{person}{Sara Solla}.} \bibinfo{year}{1989}\natexlab{}.
\newblock \showarticletitle{Optimal brain damage}.
\newblock \bibinfo{journal}{\emph{Advances in neural information processing systems}}  \bibinfo{volume}{2} (\bibinfo{year}{1989}).
\newblock


\bibitem[Li et~al\mbox{.}(2023)]%
        {li-etal-2023-compressing}
\bibfield{author}{\bibinfo{person}{Yucheng Li}, \bibinfo{person}{Bo Dong}, \bibinfo{person}{Frank Guerin}, {and} \bibinfo{person}{Chenghua Lin}.} \bibinfo{year}{2023}\natexlab{}.
\newblock \showarticletitle{Compressing Context to Enhance Inference Efficiency of Large Language Models}. In \bibinfo{booktitle}{\emph{Proceedings of the 2023 Conference on Empirical Methods in Natural Language Processing}}, \bibfield{editor}{\bibinfo{person}{Houda Bouamor}, \bibinfo{person}{Juan Pino}, {and} \bibinfo{person}{Kalika Bali}} (Eds.). \bibinfo{publisher}{Association for Computational Linguistics}, \bibinfo{address}{Singapore}, \bibinfo{pages}{6342--6353}.
\newblock
\href{https://doi.org/10.18653/v1/2023.emnlp-main.391}{doi:\nolinkurl{10.18653/v1/2023.emnlp-main.391}}


\bibitem[Li et~al\mbox{.}(2024)]%
        {li2024promptcompressionlargelanguage}
\bibfield{author}{\bibinfo{person}{Zongqian Li}, \bibinfo{person}{Yinhong Liu}, \bibinfo{person}{Yixuan Su}, {and} \bibinfo{person}{Nigel Collier}.} \bibinfo{year}{2024}\natexlab{}.
\newblock \bibinfo{title}{Prompt Compression for Large Language Models: A Survey}.
\newblock
\showeprint[arxiv]{2410.12388}~[cs.CL]
\urldef\tempurl%
\url{https://arxiv.org/abs/2410.12388}
\showURL{%
\tempurl}


\bibitem[Lin et~al\mbox{.}(2021)]%
        {pyserini}
\bibfield{author}{\bibinfo{person}{Jimmy Lin}, \bibinfo{person}{Xueguang Ma}, \bibinfo{person}{Sheng-Chieh Lin}, \bibinfo{person}{Jheng-Hong Yang}, \bibinfo{person}{Ronak Pradeep}, {and} \bibinfo{person}{Rodrigo Nogueira}.} \bibinfo{year}{2021}\natexlab{}.
\newblock \showarticletitle{Pyserini: A Python Toolkit for Reproducible Information Retrieval Research with Sparse and Dense Representations}. In \bibinfo{booktitle}{\emph{Proceedings of the 44th International ACM SIGIR Conference on Research and Development in Information Retrieval}} (Virtual Event, Canada) \emph{(\bibinfo{series}{SIGIR '21})}. \bibinfo{publisher}{Association for Computing Machinery}, \bibinfo{address}{New York, NY, USA}, \bibinfo{pages}{2356–2362}.
\newblock
\showISBNx{9781450380379}
\href{https://doi.org/10.1145/3404835.3463238}{doi:\nolinkurl{10.1145/3404835.3463238}}


\bibitem[Liu et~al\mbox{.}(2025)]%
        {Zhengmrlreranker}
\bibfield{author}{\bibinfo{person}{Zheng Liu}, \bibinfo{person}{Chaofan Li}, \bibinfo{person}{Shitao Xiao}, \bibinfo{person}{Chaozhuo Li}, \bibinfo{person}{Chen~Jason Zhang}, \bibinfo{person}{Hao Liao}, \bibinfo{person}{Defu Lian}, {and} \bibinfo{person}{Yingxia Shao}.} \bibinfo{year}{2025}\natexlab{}.
\newblock \showarticletitle{Fitting Into Any Shape: A Flexible LLM-Based Re-Ranker With Configurable Depth and Width}. In \bibinfo{booktitle}{\emph{Proceedings of the ACM on Web Conference 2025}} (Sydney NSW, Australia) \emph{(\bibinfo{series}{WWW '25})}. \bibinfo{publisher}{Association for Computing Machinery}, \bibinfo{address}{New York, NY, USA}, \bibinfo{pages}{3942–3951}.
\newblock
\showISBNx{9798400712746}
\href{https://doi.org/10.1145/3696410.3714620}{doi:\nolinkurl{10.1145/3696410.3714620}}


\bibitem[Liu et~al\mbox{.}(2024)]%
        {liu2024kiviatuningfree}
\bibfield{author}{\bibinfo{person}{Zirui Liu}, \bibinfo{person}{Jiayi Yuan}, \bibinfo{person}{Hongye Jin}, \bibinfo{person}{Shaochen Zhong}, \bibinfo{person}{Zhaozhuo Xu}, \bibinfo{person}{Vladimir Braverman}, \bibinfo{person}{Beidi Chen}, {and} \bibinfo{person}{Xia Hu}.} \bibinfo{year}{2024}\natexlab{}.
\newblock \showarticletitle{{KIVI}: A Tuning-Free Asymmetric 2bit Quantization for {KV} Cache}. In \bibinfo{booktitle}{\emph{Proceedings of the 41st International Conference on Machine Learning}} \emph{(\bibinfo{series}{Proceedings of Machine Learning Research}, Vol.~\bibinfo{volume}{235})}, \bibfield{editor}{\bibinfo{person}{Ruslan Salakhutdinov}, \bibinfo{person}{Zico Kolter}, \bibinfo{person}{Katherine Heller}, \bibinfo{person}{Adrian Weller}, \bibinfo{person}{Nuria Oliver}, \bibinfo{person}{Jonathan Scarlett}, {and} \bibinfo{person}{Felix Berkenkamp}} (Eds.). \bibinfo{publisher}{PMLR}, \bibinfo{pages}{32332--32344}.
\newblock
\urldef\tempurl%
\url{https://proceedings.mlr.press/v235/liu24bz.html}
\showURL{%
\tempurl}


\bibitem[Ma et~al\mbox{.}(2025)]%
        {tevatronv2}
\bibfield{author}{\bibinfo{person}{Xueguang Ma}, \bibinfo{person}{Luyu Gao}, \bibinfo{person}{Shengyao Zhuang}, \bibinfo{person}{Jiaqi~Samantha Zhan}, \bibinfo{person}{Jamie Callan}, {and} \bibinfo{person}{Jimmy Lin}.} \bibinfo{year}{2025}\natexlab{}.
\newblock \showarticletitle{Tevatron 2.0: Unified Document Retrieval Toolkit across Scale, Language, and Modality}. In \bibinfo{booktitle}{\emph{Proceedings of the 48th International ACM SIGIR Conference on Research and Development in Information Retrieval}} (Padua, Italy) \emph{(\bibinfo{series}{SIGIR '25})}. \bibinfo{publisher}{Association for Computing Machinery}, \bibinfo{address}{New York, NY, USA}, \bibinfo{pages}{4061–4065}.
\newblock
\showISBNx{9798400715921}
\href{https://doi.org/10.1145/3726302.3730135}{doi:\nolinkurl{10.1145/3726302.3730135}}


\bibitem[Ma et~al\mbox{.}(2023)]%
        {ma2023zero}
\bibfield{author}{\bibinfo{person}{Xueguang Ma}, \bibinfo{person}{Xinyu Zhang}, \bibinfo{person}{Ronak Pradeep}, {and} \bibinfo{person}{Jimmy Lin}.} \bibinfo{year}{2023}\natexlab{}.
\newblock \bibinfo{title}{Zero-Shot Listwise Document Reranking with a Large Language Model}.
\newblock
\showeprint[arxiv]{2305.02156}~[cs.IR]
\urldef\tempurl%
\url{https://arxiv.org/abs/2305.02156}
\showURL{%
\tempurl}


\bibitem[Michel et~al\mbox{.}(2019)]%
        {NEURIPS2019_2c601ad9}
\bibfield{author}{\bibinfo{person}{Paul Michel}, \bibinfo{person}{Omer Levy}, {and} \bibinfo{person}{Graham Neubig}.} \bibinfo{year}{2019}\natexlab{}.
\newblock \showarticletitle{Are Sixteen Heads Really Better than One?}. In \bibinfo{booktitle}{\emph{Advances in Neural Information Processing Systems}}, \bibfield{editor}{\bibinfo{person}{H.~Wallach}, \bibinfo{person}{H.~Larochelle}, \bibinfo{person}{A.~Beygelzimer}, \bibinfo{person}{F.~d\textquotesingle Alch\'{e}-Buc}, \bibinfo{person}{E.~Fox}, {and} \bibinfo{person}{R.~Garnett}} (Eds.), Vol.~\bibinfo{volume}{32}. \bibinfo{publisher}{Curran Associates, Inc.}
\newblock
\urldef\tempurl%
\url{https://proceedings.neurips.cc/paper_files/paper/2019/file/2c601ad9d2ff9bc8b282670cdd54f69f-Paper.pdf}
\showURL{%
\tempurl}


\bibitem[Mu et~al\mbox{.}(2023)]%
        {mu2023learningtocompresspromptswithgisttokens}
\bibfield{author}{\bibinfo{person}{Jesse Mu}, \bibinfo{person}{Xiang Li}, {and} \bibinfo{person}{Noah Goodman}.} \bibinfo{year}{2023}\natexlab{}.
\newblock \showarticletitle{Learning to compress prompts with gist tokens}.
\newblock \bibinfo{journal}{\emph{Advances in Neural Information Processing Systems}}  \bibinfo{volume}{36} (\bibinfo{year}{2023}), \bibinfo{pages}{19327--19352}.
\newblock


\bibitem[Nogueira and Cho(2019)]%
        {nogueira2019passage}
\bibfield{author}{\bibinfo{person}{Rodrigo Nogueira} {and} \bibinfo{person}{Kyunghyun Cho}.} \bibinfo{year}{2019}\natexlab{}.
\newblock \showarticletitle{Passage Re-ranking with {BERT}}.
\newblock \bibinfo{journal}{\emph{CoRR}}  \bibinfo{volume}{abs/1901.04085} (\bibinfo{year}{2019}).
\newblock
\showeprint[arXiv]{1901.04085}
\urldef\tempurl%
\url{http://arxiv.org/abs/1901.04085}
\showURL{%
\tempurl}


\bibitem[Pradeep et~al\mbox{.}(2023a)]%
        {pradeep2023rankllm}
\bibfield{author}{\bibinfo{person}{Ronak Pradeep}, \bibinfo{person}{Sahel Sharifymoghaddam}, {and} \bibinfo{person}{Jimmy Lin}.} \bibinfo{year}{2023}\natexlab{a}.
\newblock \bibinfo{title}{RankVicuna: Zero-Shot Listwise Document Reranking with Open-Source Large Language Models}.
\newblock
\showeprint[arxiv]{2309.15088}~[cs.IR]
\urldef\tempurl%
\url{https://arxiv.org/abs/2309.15088}
\showURL{%
\tempurl}


\bibitem[Pradeep et~al\mbox{.}(2023b)]%
        {RankZephyr}
\bibfield{author}{\bibinfo{person}{Ronak Pradeep}, \bibinfo{person}{Sahel Sharifymoghaddam}, {and} \bibinfo{person}{Jimmy Lin}.} \bibinfo{year}{2023}\natexlab{b}.
\newblock \bibinfo{title}{RankZephyr: Effective and Robust Zero-Shot Listwise Reranking is a Breeze!}
\newblock
\showeprint[arxiv]{2312.02724}~[cs.IR]
\urldef\tempurl%
\url{https://arxiv.org/abs/2312.02724}
\showURL{%
\tempurl}


\bibitem[Sanh et~al\mbox{.}(2020)]%
        {sanh2020distilbertdistilledversionbert}
\bibfield{author}{\bibinfo{person}{Victor Sanh}, \bibinfo{person}{Lysandre Debut}, \bibinfo{person}{Julien Chaumond}, {and} \bibinfo{person}{Thomas Wolf}.} \bibinfo{year}{2020}\natexlab{}.
\newblock \bibinfo{title}{DistilBERT, a distilled version of BERT: smaller, faster, cheaper and lighter}.
\newblock
\showeprint[arxiv]{1910.01108}~[cs.CL]
\urldef\tempurl%
\url{https://arxiv.org/abs/1910.01108}
\showURL{%
\tempurl}


\bibitem[Sun et~al\mbox{.}(2023)]%
        {sun2023chatgpt}
\bibfield{author}{\bibinfo{person}{Weiwei Sun}, \bibinfo{person}{Lingyong Yan}, \bibinfo{person}{Xinyu Ma}, \bibinfo{person}{Shuaiqiang Wang}, \bibinfo{person}{Pengjie Ren}, \bibinfo{person}{Zhumin Chen}, \bibinfo{person}{Dawei Yin}, {and} \bibinfo{person}{Zhaochun Ren}.} \bibinfo{year}{2023}\natexlab{}.
\newblock \showarticletitle{Is {C}hat{GPT} Good at Search? Investigating Large Language Models as Re-Ranking Agents}. In \bibinfo{booktitle}{\emph{Proceedings of the 2023 Conference on Empirical Methods in Natural Language Processing}}, \bibfield{editor}{\bibinfo{person}{Houda Bouamor}, \bibinfo{person}{Juan Pino}, {and} \bibinfo{person}{Kalika Bali}} (Eds.). \bibinfo{publisher}{Association for Computational Linguistics}, \bibinfo{address}{Singapore}, \bibinfo{pages}{14918--14937}.
\newblock
\href{https://doi.org/10.18653/v1/2023.emnlp-main.923}{doi:\nolinkurl{10.18653/v1/2023.emnlp-main.923}}


\bibitem[Wei et~al\mbox{.}(2025)]%
        {wei2025deepseek}
\bibfield{author}{\bibinfo{person}{Haoran Wei}, \bibinfo{person}{Yaofeng Sun}, {and} \bibinfo{person}{Yukun Li}.} \bibinfo{year}{2025}\natexlab{}.
\newblock \showarticletitle{DeepSeek-OCR: Contexts Optical Compression}.
\newblock \bibinfo{journal}{\emph{arXiv preprint arXiv:2510.18234}} (\bibinfo{year}{2025}).
\newblock


\bibitem[Weller et~al\mbox{.}(2025)]%
        {weller2025rank1testtimecomputereranking}
\bibfield{author}{\bibinfo{person}{Orion Weller}, \bibinfo{person}{Kathryn Ricci}, \bibinfo{person}{Eugene Yang}, \bibinfo{person}{Andrew Yates}, \bibinfo{person}{Dawn Lawrie}, {and} \bibinfo{person}{Benjamin~Van Durme}.} \bibinfo{year}{2025}\natexlab{}.
\newblock \bibinfo{title}{Rank1: Test-Time Compute for Reranking in Information Retrieval}.
\newblock
\showeprint[arxiv]{2502.18418}~[cs.IR]
\urldef\tempurl%
\url{https://arxiv.org/abs/2502.18418}
\showURL{%
\tempurl}


\bibitem[Xia et~al\mbox{.}(2024)]%
        {xia2024shearedllama}
\bibfield{author}{\bibinfo{person}{Mengzhou Xia}, \bibinfo{person}{Tianyu Gao}, \bibinfo{person}{Zhiyuan Zeng}, {and} \bibinfo{person}{Danqi Chen}.} \bibinfo{year}{2024}\natexlab{}.
\newblock \showarticletitle{Sheared {LL}a{MA}: Accelerating Language Model Pre-training via Structured Pruning}. In \bibinfo{booktitle}{\emph{The Twelfth International Conference on Learning Representations}}.
\newblock
\urldef\tempurl%
\url{https://openreview.net/forum?id=09iOdaeOzp}
\showURL{%
\tempurl}


\bibitem[Xiao et~al\mbox{.}(2024)]%
        {xiao2024efficientstreaming}
\bibfield{author}{\bibinfo{person}{Guangxuan Xiao}, \bibinfo{person}{Yuandong Tian}, \bibinfo{person}{Beidi Chen}, \bibinfo{person}{Song Han}, {and} \bibinfo{person}{Mike Lewis}.} \bibinfo{year}{2024}\natexlab{}.
\newblock \showarticletitle{Efficient Streaming Language Models with Attention Sinks}. In \bibinfo{booktitle}{\emph{The Twelfth International Conference on Learning Representations}}.
\newblock
\urldef\tempurl%
\url{https://openreview.net/forum?id=NG7sS51zVF}
\showURL{%
\tempurl}


\bibitem[Xu(2026)]%
        {xu2026rankmambabenchmarkingmambasdocument}
\bibfield{author}{\bibinfo{person}{Zhichao Xu}.} \bibinfo{year}{2026}\natexlab{}.
\newblock \bibinfo{title}{RankMamba: Benchmarking Mamba's Document Ranking Performance in the Era of Transformers}.
\newblock
\showeprint[arxiv]{2403.18276}~[cs.IR]
\urldef\tempurl%
\url{https://arxiv.org/abs/2403.18276}
\showURL{%
\tempurl}


\bibitem[Xu et~al\mbox{.}(2024)]%
        {xu-etal-2024-beyond-perplexity}
\bibfield{author}{\bibinfo{person}{Zhichao Xu}, \bibinfo{person}{Ashim Gupta}, \bibinfo{person}{Tao Li}, \bibinfo{person}{Oliver Bentham}, {and} \bibinfo{person}{Vivek Srikumar}.} \bibinfo{year}{2024}\natexlab{}.
\newblock \showarticletitle{Beyond Perplexity: Multi-dimensional Safety Evaluation of {LLM} Compression}. In \bibinfo{booktitle}{\emph{Findings of the Association for Computational Linguistics: EMNLP 2024}}, \bibfield{editor}{\bibinfo{person}{Yaser Al-Onaizan}, \bibinfo{person}{Mohit Bansal}, {and} \bibinfo{person}{Yun-Nung Chen}} (Eds.). \bibinfo{publisher}{Association for Computational Linguistics}, \bibinfo{address}{Miami, Florida, USA}, \bibinfo{pages}{15359--15396}.
\newblock
\href{https://doi.org/10.18653/v1/2024.findings-emnlp.901}{doi:\nolinkurl{10.18653/v1/2024.findings-emnlp.901}}


\bibitem[Xu et~al\mbox{.}(2025a)]%
        {xu-etal-2025-distillation}
\bibfield{author}{\bibinfo{person}{Zhichao Xu}, \bibinfo{person}{Zhiqi Huang}, \bibinfo{person}{Shengyao Zhuang}, {and} \bibinfo{person}{Vivek Srikumar}.} \bibinfo{year}{2025}\natexlab{a}.
\newblock \showarticletitle{Distillation versus Contrastive Learning: How to Train Your Rerankers}. In \bibinfo{booktitle}{\emph{Proceedings of the 14th International Joint Conference on Natural Language Processing and the 4th Conference of the Asia-Pacific Chapter of the Association for Computational Linguistics}}, \bibfield{editor}{\bibinfo{person}{Kentaro Inui}, \bibinfo{person}{Sakriani Sakti}, \bibinfo{person}{Haofen Wang}, \bibinfo{person}{Derek~F. Wong}, \bibinfo{person}{Pushpak Bhattacharyya}, \bibinfo{person}{Biplab Banerjee}, \bibinfo{person}{Asif Ekbal}, \bibinfo{person}{Tanmoy Chakraborty}, {and} \bibinfo{person}{Dhirendra~Pratap Singh}} (Eds.). \bibinfo{publisher}{The Asian Federation of Natural Language Processing and The Association for Computational Linguistics}, \bibinfo{address}{Mumbai, India}, \bibinfo{pages}{564--578}.
\newblock
\showISBNx{979-8-89176-303-6}
\urldef\tempurl%
\url{https://aclanthology.org/2025.findings-ijcnlp.33/}
\showURL{%
\tempurl}


\bibitem[Xu et~al\mbox{.}(2025b)]%
        {xu2025surveymodelarchitecturesinformation}
\bibfield{author}{\bibinfo{person}{Zhichao Xu}, \bibinfo{person}{Fengran Mo}, \bibinfo{person}{Zhiqi Huang}, \bibinfo{person}{Crystina Zhang}, \bibinfo{person}{Puxuan Yu}, \bibinfo{person}{Bei Wang}, \bibinfo{person}{Jimmy Lin}, {and} \bibinfo{person}{Vivek Srikumar}.} \bibinfo{year}{2025}\natexlab{b}.
\newblock \bibinfo{title}{A Survey of Model Architectures in Information Retrieval}.
\newblock
\showeprint[arxiv]{2502.14822}~[cs.IR]
\urldef\tempurl%
\url{https://arxiv.org/abs/2502.14822}
\showURL{%
\tempurl}


\bibitem[Yang et~al\mbox{.}(2025a)]%
        {qwen3}
\bibfield{author}{\bibinfo{person}{An Yang}, \bibinfo{person}{Anfeng Li}, \bibinfo{person}{Baosong Yang}, \bibinfo{person}{Beichen Zhang}, \bibinfo{person}{Binyuan Hui}, \bibinfo{person}{Bo Zheng}, \bibinfo{person}{Bowen Yu}, \bibinfo{person}{Chang Gao}, \bibinfo{person}{Chengen Huang}, \bibinfo{person}{Chenxu Lv}, \bibinfo{person}{Chujie Zheng}, \bibinfo{person}{Dayiheng Liu}, \bibinfo{person}{Fan Zhou}, \bibinfo{person}{Fei Huang}, \bibinfo{person}{Feng Hu}, \bibinfo{person}{Hao Ge}, \bibinfo{person}{Haoran Wei}, \bibinfo{person}{Huan Lin}, \bibinfo{person}{Jialong Tang}, \bibinfo{person}{Jian Yang}, \bibinfo{person}{Jianhong Tu}, \bibinfo{person}{Jianwei Zhang}, \bibinfo{person}{Jianxin Yang}, \bibinfo{person}{Jiaxi Yang}, \bibinfo{person}{Jing Zhou}, \bibinfo{person}{Jingren Zhou}, \bibinfo{person}{Junyang Lin}, \bibinfo{person}{Kai Dang}, \bibinfo{person}{Keqin Bao}, \bibinfo{person}{Kexin Yang}, \bibinfo{person}{Le Yu}, \bibinfo{person}{Lianghao Deng}, \bibinfo{person}{Mei Li}, \bibinfo{person}{Mingfeng
  Xue}, \bibinfo{person}{Mingze Li}, \bibinfo{person}{Pei Zhang}, \bibinfo{person}{Peng Wang}, \bibinfo{person}{Qin Zhu}, \bibinfo{person}{Rui Men}, \bibinfo{person}{Ruize Gao}, \bibinfo{person}{Shixuan Liu}, \bibinfo{person}{Shuang Luo}, \bibinfo{person}{Tianhao Li}, \bibinfo{person}{Tianyi Tang}, \bibinfo{person}{Wenbiao Yin}, \bibinfo{person}{Xingzhang Ren}, \bibinfo{person}{Xinyu Wang}, \bibinfo{person}{Xinyu Zhang}, \bibinfo{person}{Xuancheng Ren}, \bibinfo{person}{Yang Fan}, \bibinfo{person}{Yang Su}, \bibinfo{person}{Yichang Zhang}, \bibinfo{person}{Yinger Zhang}, \bibinfo{person}{Yu Wan}, \bibinfo{person}{Yuqiong Liu}, \bibinfo{person}{Zekun Wang}, \bibinfo{person}{Zeyu Cui}, \bibinfo{person}{Zhenru Zhang}, \bibinfo{person}{Zhipeng Zhou}, {and} \bibinfo{person}{Zihan Qiu}.} \bibinfo{year}{2025}\natexlab{a}.
\newblock \bibinfo{title}{Qwen3 Technical Report}.
\newblock
\showeprint[arxiv]{2505.09388}~[cs.CL]
\urldef\tempurl%
\url{https://arxiv.org/abs/2505.09388}
\showURL{%
\tempurl}


\bibitem[Yang et~al\mbox{.}(2025b)]%
        {yang2025rankktesttimereasoninglistwise}
\bibfield{author}{\bibinfo{person}{Eugene Yang}, \bibinfo{person}{Andrew Yates}, \bibinfo{person}{Kathryn Ricci}, \bibinfo{person}{Orion Weller}, \bibinfo{person}{Vivek Chari}, \bibinfo{person}{Benjamin~Van Durme}, {and} \bibinfo{person}{Dawn Lawrie}.} \bibinfo{year}{2025}\natexlab{b}.
\newblock \bibinfo{title}{Rank-K: Test-Time Reasoning for Listwise Reranking}.
\newblock
\showeprint[arxiv]{2505.14432}~[cs.IR]
\urldef\tempurl%
\url{https://arxiv.org/abs/2505.14432}
\showURL{%
\tempurl}


\bibitem[Yates et~al\mbox{.}(2021)]%
        {bert2021yates}
\bibfield{author}{\bibinfo{person}{Andrew Yates}, \bibinfo{person}{Rodrigo Nogueira}, {and} \bibinfo{person}{Jimmy Lin}.} \bibinfo{year}{2021}\natexlab{}.
\newblock \showarticletitle{Pretrained Transformers for Text Ranking: BERT and Beyond}. In \bibinfo{booktitle}{\emph{Proceedings of the 44th International ACM SIGIR Conference on Research and Development in Information Retrieval}} (Virtual Event, Canada) \emph{(\bibinfo{series}{SIGIR '21})}. \bibinfo{publisher}{Association for Computing Machinery}, \bibinfo{address}{New York, NY, USA}, \bibinfo{pages}{2666–2668}.
\newblock
\showISBNx{9781450380379}
\href{https://doi.org/10.1145/3404835.3462812}{doi:\nolinkurl{10.1145/3404835.3462812}}


\bibitem[Zafrir et~al\mbox{.}(2019)]%
        {Zafrir_2019}
\bibfield{author}{\bibinfo{person}{Ofir Zafrir}, \bibinfo{person}{Guy Boudoukh}, \bibinfo{person}{Peter Izsak}, {and} \bibinfo{person}{Moshe Wasserblat}.} \bibinfo{year}{2019}\natexlab{}.
\newblock \showarticletitle{Q8BERT: Quantized 8Bit BERT}. In \bibinfo{booktitle}{\emph{2019 Fifth Workshop on Energy Efficient Machine Learning and Cognitive Computing - NeurIPS Edition (EMC2-NIPS)}}. \bibinfo{publisher}{IEEE}, \bibinfo{pages}{36–39}.
\newblock
\href{https://doi.org/10.1109/emc2-nips53020.2019.00016}{doi:\nolinkurl{10.1109/emc2-nips53020.2019.00016}}


\bibitem[Zhang et~al\mbox{.}(2025)]%
        {zhang2025jasper}
\bibfield{author}{\bibinfo{person}{Dun Zhang}, \bibinfo{person}{Ziyang Zeng}, \bibinfo{person}{Yudong Zhou}, {and} \bibinfo{person}{Shuyang Lu}.} \bibinfo{year}{2025}\natexlab{}.
\newblock \bibinfo{title}{Jasper-Token-Compression-600M Technical Report}.
\newblock
\showeprint[arxiv]{2511.14405}~[cs.IR]
\urldef\tempurl%
\url{https://arxiv.org/abs/2511.14405}
\showURL{%
\tempurl}


\bibitem[Zhuang et~al\mbox{.}(2025)]%
        {zhuang2025rankr1enhancingreasoningllmbased}
\bibfield{author}{\bibinfo{person}{Shengyao Zhuang}, \bibinfo{person}{Xueguang Ma}, \bibinfo{person}{Bevan Koopman}, \bibinfo{person}{Jimmy Lin}, {and} \bibinfo{person}{Guido Zuccon}.} \bibinfo{year}{2025}\natexlab{}.
\newblock \bibinfo{title}{Rank-R1: Enhancing Reasoning in LLM-based Document Rerankers via Reinforcement Learning}.
\newblock
\showeprint[arxiv]{2503.06034}~[cs.IR]
\urldef\tempurl%
\url{https://arxiv.org/abs/2503.06034}
\showURL{%
\tempurl}


\bibitem[Zhuang et~al\mbox{.}(2024)]%
        {setwise}
\bibfield{author}{\bibinfo{person}{Shengyao Zhuang}, \bibinfo{person}{Honglei Zhuang}, \bibinfo{person}{Bevan Koopman}, {and} \bibinfo{person}{Guido Zuccon}.} \bibinfo{year}{2024}\natexlab{}.
\newblock \showarticletitle{A Setwise Approach for Effective and Highly Efficient Zero-shot Ranking with Large Language Models}. In \bibinfo{booktitle}{\emph{Proceedings of the 47th International ACM SIGIR Conference on Research and Development in Information Retrieval}} (Washington DC, USA) \emph{(\bibinfo{series}{SIGIR '24})}. \bibinfo{publisher}{Association for Computing Machinery}, \bibinfo{address}{New York, NY, USA}, \bibinfo{pages}{38–47}.
\newblock
\showISBNx{9798400704314}
\href{https://doi.org/10.1145/3626772.3657813}{doi:\nolinkurl{10.1145/3626772.3657813}}


\bibitem[Zhuang and Zuccon(2021)]%
        {zhuang2021fastpassagererankingcontextualized}
\bibfield{author}{\bibinfo{person}{Shengyao Zhuang} {and} \bibinfo{person}{Guido Zuccon}.} \bibinfo{year}{2021}\natexlab{}.
\newblock \bibinfo{title}{Fast Passage Re-ranking with Contextualized Exact Term Matching and Efficient Passage Expansion}.
\newblock
\showeprint[arxiv]{2108.08513}~[cs.IR]
\urldef\tempurl%
\url{https://arxiv.org/abs/2108.08513}
\showURL{%
\tempurl}


\end{thebibliography}

\end{document}